\documentclass{article}
\usepackage{spconf,amsmath,graphicx}
\usepackage{multirow, amsfonts, makecell}
\usepackage{xcolor}
\usepackage{pythonhighlight}
\usepackage[belowskip=-5pt,aboveskip=0pt]{caption}

\title{TorchAudio-Squim: Reference-less Speech Quality and Intelligibility measures in TorchAudio}
\name{Anurag Kumar$^1$, Ke Tan$^1$, Zhaoheng Ni$^2$, Pranay Manocha$^{3*}$\thanks{*Work done during internship at Meta.}, Xiaohui Zhang$^2$, Ethan Henderson$^1$, Buye Xu$^1$}
\address{$^1$ Meta Reality Labs Research, $^2$ Meta, $^3$ Princeton University}
\begin{document}
\ninept
\maketitle
\begin{abstract}
Measuring quality and intelligibility of a speech signal is usually a critical step in development of speech processing systems. To enable this, a variety of metrics to measure quality and intelligibility under different assumptions have been developed. Through this paper, we introduce tools and a set of models to estimate such known metrics using deep neural networks. These models are made available in the well-established \emph{TorchAudio} library, the core audio and speech processing library within the PyTorch deep learning framework. We refer to it as \emph{TorchAudio-Squim}, TorchAudio-Speech QUality and Intelligibility Measures.  More specifically, in the current version of TorchAudio-squim, we establish and release models for estimating PESQ, STOI and SI-SDR among objective metrics and MOS among subjective metrics. We develop a \emph{novel} approach for objective metric estimation and use a recently developed approach for subjective metric estimation. These models operate in a ``reference-less" manner, that is they do not require the corresponding clean speech as reference for speech assessment. Given the unavailability of clean speech and the effortful process of subjective evaluation in real-world situations, such easy-to-use tools would greatly benefit speech processing research and development.
\end{abstract}
\section{Introduction}
\label{sec:intro}
Speech processing systems more often than not deal with degraded or corrupted speech signals. 
Hence, in design and development of such systems, there is a need to measure the quality of speech signals. Over the years, a variety of methods/metrics have been developed for speech assessment. While these methods measure degradations in the speech signals under certain assumptions, for the purposes of this paper we consider two broad classes of metrics: \emph{subjective metrics}, which are obtained through human listening tests of the speech signals, and \emph{objective metrics}, which do not require human judgements and are derived primarily by comparing the given speech signal (degraded) to the corresponding reference clean speech. This reliance on reference clean speech makes these metrics ``intrusive" as opposed to ``non-intrusive" methods which do not require reference clean speech. Note that, there are certain objective metrics which may not require reference clean speech \cite{loizou2011speech, gray2000non}, but for the purposes of this paper objective metrics would refer to the more common cases where reference clean speech is required \cite{loizou2011speech}.

Both subjective and objective metrics have their own merits and disadvantages. Compared to objective methods, subjective methods are the more reliable approach to assess speech signals as they are based on human perceptions and judgements. However, human-based evaluations are not scalable and often difficult to do. They require expert listeners in most situations and can be a time-consuming and tedious process. Objective metrics avoid these problems but come with their own constraints. They may not always correlate well with subjective assessments. More importantly though, they require the reference clean speech to assess quality or intelligibility, making them impractical for real-world uses, where the reference clean signals are usually unavailable. 

To address the above constraints around both subjective and objective metrics, in recent years there have been efforts on building machine learning based estimators of these metrics. On the subjective metrics front, recently a challenge on Mean Opinion Score (MOS) prediction was organized~\cite{huang2022voicemos}. Several works have explored training neural networks for speech MOS estimation~\cite{lo2019mosnet,patton2016automos,mittag2021nisqa,cooper2021generalization,huang2021ldnet,reddy2021dnsmos,catellier2020wawenets,zhang2021end}. Some such as DNSMOS~\cite{reddy2022dnsmos} are trained on large-scale crowdsourced data of MOS ratings for speech, while others leverage pre-trained self-supervised models for improved estimation~\cite{cooper2021generalization}. On the objective metric front also, there have been quite a few works on reference-less estimation of well-known objective metrics such as PESQ, STOI, ESTOI, HASQI, etc.~\cite{dong2020attention, dong2019classification, zezario2020stoi, yu2021metricnet, zezario2022deep}. 

While these research works have led to progress in development of metric estimators, widespread use of these reference-less estimators in different speech applications is still uncommon. The primary reason is lack of simple and easy-to-use tools and inference models which can be readily integrated into existing speech systems. We aim to address this problem through this paper. Moreover, such open-source tools and inference models would augment and support future research works on metric estimation as well. 

We keep the following criteria and principles in mind for our system. \textbf{\emph{(1)}} Usable in real-world applications with ease. Since reference clean speech is often unavailable, we only focus on developing \emph{non-intrusive methods} which can estimate metrics without reference clean speech. \textbf{\emph{(2)}} Deep learning plays a critical role in the development of a large number of current speech systems, and thus our system will primarily be deep neural network based as well. This also ensures that we have differentiable estimators which can be easily utilized for training other deep learning based speech systems. \textbf{\emph{(3)}} Both subjective and objective metrics are widely used, and hence estimators of metrics falling in both classes will be part of the system. \textbf{\emph{(4)}} The tools and models will be continuously developed, and better models will be released regularly. This includes updating the models for better metric estimation, reducing computational load, extending the models to estimate more objective and subjective metrics. 

To this end, we are releasing \emph{TorchAudio-Squim} within the \emph{TorchAudio}~\cite{yang2022TorchAudio} library for estimating speech assessment metrics. \emph{TorchAudio} is the official audio domain library of PyTorch, which supports essential building blocks of audio and speech processing and enables advancement of research in various audio and speech problems. By integrating speech quality and intelligibility assessment components into TorchAudio, our goal is to ease the use of these metrics in design and development of speech processing systems. The current version of \emph{TorchAudio-Squim} enables estimation of 3 objective metrics and 1 subjective metric.

\section{TorchAudio-Squim Overview}
\label{sec:format}

We give a quick overview of the speech quality and intelligibility assessment tools currently provided through \emph{TorchAudio-Squim}. As mentioned before, we are supporting both subjective and objective metrics in \emph{TorchAudio-Squim}. 

\textbf{Objective Metrics}: In the current version, we are releasing a model to estimate 3 well-known objective metrics for speech assessment. These are Perceptual Evaluation of Speech Quality (PESQ)~\cite{rix2001perceptual, rec2005p},   Short-Time Objective Intelligibility (STOI)~\cite{taal2011algorithm} and Scale-Invariant Signal-to-Distortion Ratio (SI-SDR)~\cite{le2019sdr}. Note that, we use Wideband-PESQ~\cite{rec2005p} and the term PESQ will refer to WB-PESQ throughout this paper. In this work, we develop a single network to estimate these 3 metrics. Our approach to objective metric estimation is \emph{novel}, and we present complete details of the approach along with comprehensive experimental results and analyses in further sections. 

\textbf{Subjective Metrics}: On the subjective metric front, we provide a model to estimate the Mean Opinion Score (MOS) rating. Mean opinion scores are ratings between 1 to 5 given by human listeners to quantify quality of a given speech signal. In this version, we are releasing the NORESQA-MOS~\cite{manocha2022speech} model to estimate MOS. This approach uses one or more random clean speech samples from a database as reference(s). More details are available in Section \ref{sec:subjective}.

Within \emph{TorchAudio}, the model architectures are defined under \emph{torchaudio.models} module, and the pre-trained models are defined under \emph{torchaudio.pipelines} which provide end-to-end solutions for speech quality and intelligibility assessment.

The following example code shows how to estimate the MOS, STOI, PESQ, and SI-SDR scores using \emph{TorchAudio} library\footnote{https://pytorch.org/audio/main/prototype.pipelines.html}:

\begin{python}
from torchaudio.prototype.pipelines import {
    SQUIM_OBJECTIVE,
    SQUIM_SUBJECTIVE
}

subjective_model = SQUIM_SUBJECTIVE.get_model()
objective_model = SQUIM_OBJECTIVE.get_model()

mos = subjective_model(
    test_waveform,
    non_match_reference,
)
stoi, pesq, si_sdr = objective_model(
    test_waveform
)
\end{python}

\vspace{-5pt}
\section{Subjective Metrics: System Description}
\label{sec:subjective}
We use NORESQA-MOS~\cite{manocha2022speech} to estimate MOS for a given speech signal. Unlike more common approaches (e.g DNSMOS~\cite{reddy2021dnsmos}, NISQA~\cite{mittag2021nisqa}) which attempt to directly estimate MOS from a given sample, NORESQA-MOS relies on the idea of using non-matching references for a more grounded estimation~\cite{manocha2021noresqa}.
More concretely, the model takes in a clean speech signal sampled from any database along with the test speech sample to predict the MOS rating for it. Note that, the use of a non-matching reference (NMR) does not impact the utility of this model compared to ``pure" reference-less approaches. Any clean speech signal can be used as NMR input. 

The MOS model released in the current version of TorchAudio-Squim is the same as the one in~\cite{manocha2022speech}, where the NORESQA-MOS was analyzed and evaluated comprehensively. We request readers to refer to this paper for details.

\section{Objective Metrics: System Description}
\label{sec:print}
\begin{figure*}
    \centering
    \includegraphics[width=0.95\textwidth]{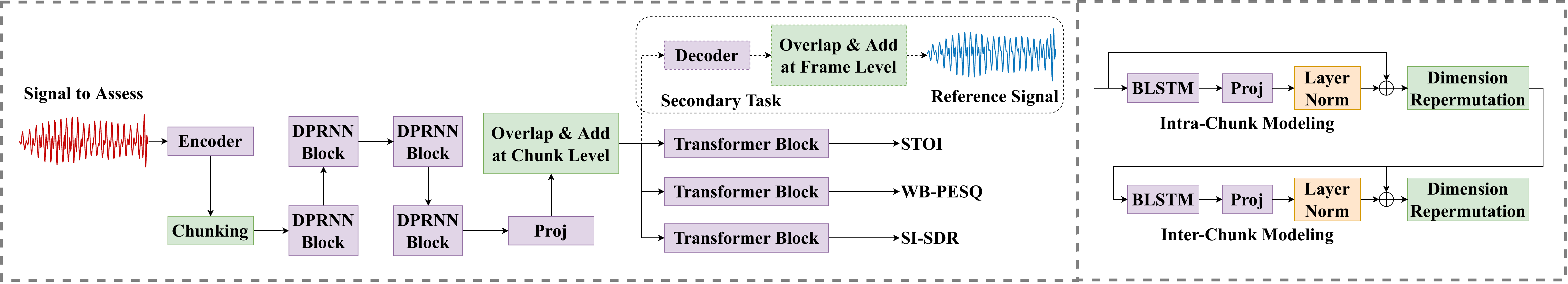}
    \caption{\textbf{Left}: Training framework for Objective Metrics
    \textbf{Right}: Details of DPRNN Block.}
    \label{fig:system_overview}
\end{figure*}

In this section, we describe details of our objective metric estimation approach. The overall schema of the approach is shown in the left panel of Fig.~\ref{fig:system_overview}, which uses a deep neural network operating on time-domain speech signals and estimates all three objective metrics in one go. We propose a novel architecture based on dual-path recurrent neural networks (DPRNNs)~\cite{luo2020dual} to perform sequential modeling of an input time-domain signal, and the learned representations are then consumed by multiple transformer based branches for metric specific estimation. Moreover, we also propose a novel multi-task training strategy for improvements in estimation of the metrics.

\subsection{Sequential modeling with dual-path RNN blocks}

\begin{figure}
    \centering
    \includegraphics[width=0.5\columnwidth]{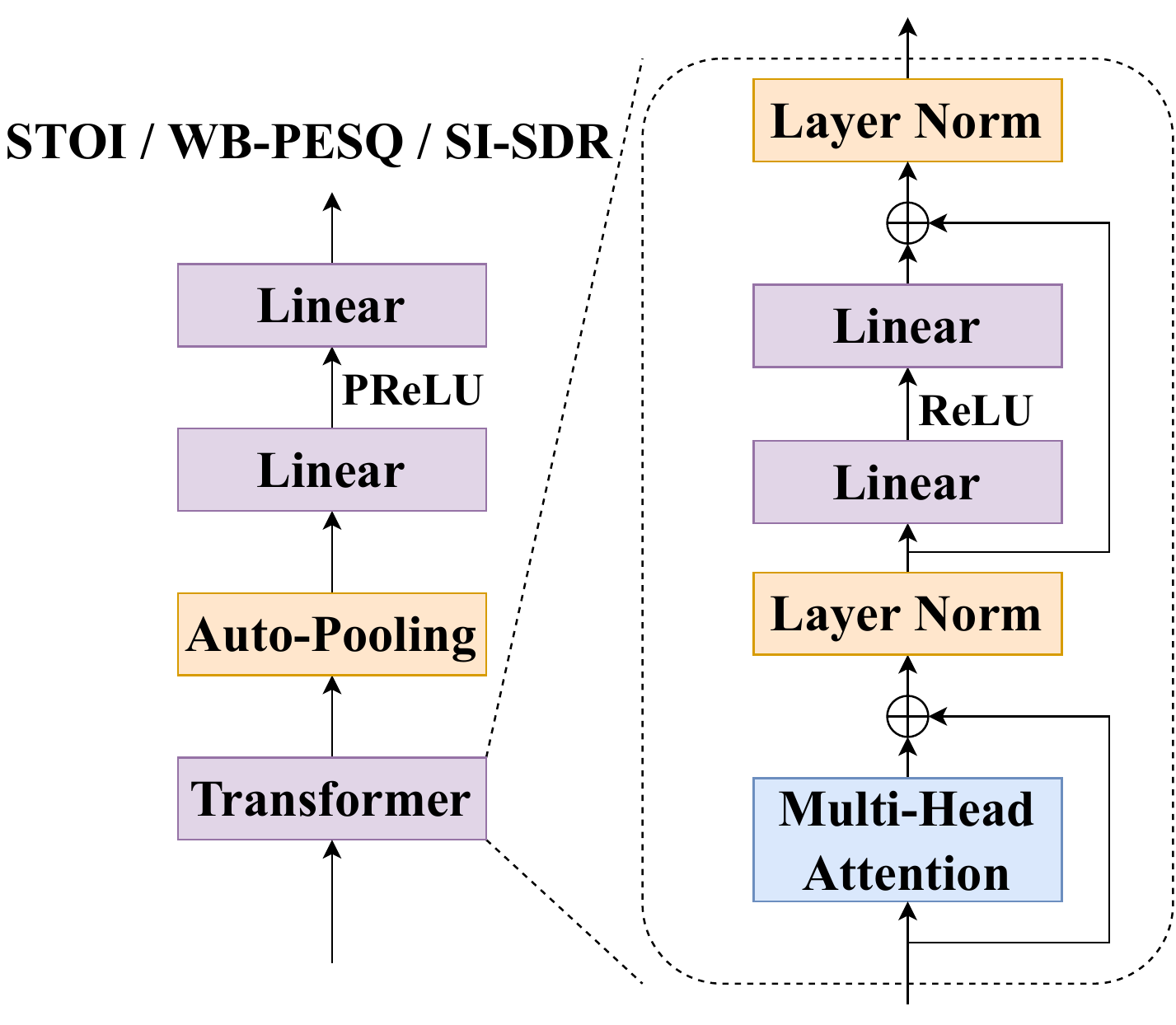}
    \caption{Details of the transformer block.}
    \label{fig:transformer_block}
\end{figure}

Given a length-$T$ signal waveform $y \in \mathbb{R}^T$ to assess, we use a strided 1-D convolutional layer followed by a rectified linear unit (ReLU) function to segment and encode it, leading to $L$ overlapped time frames of representations. With $N$ output channels, this convolutional layer has a kernel size of $P$ and a stride size of $\lfloor P / 2 \rfloor$, hence a frame size of $P$ and a hop size of $\lfloor P / 2 \rfloor$, respectively. This sequence of time frames is then divided into $S$ overlapped chunks with a chunk size of $R$ and a hop size of $\lfloor R/2 \rfloor$.

Subsequently, a stack of four DPRNN blocks is employed to process the chunk sequence $\mathbf{U} = [U_1, \dots, U_S] \in \mathbb{R}^{N \times S \times R}$, where $U_1, \dots, U_S \in \mathbb{R}^{N \times R}$. Such block processing can be written as 
\begin{equation}
\mathbf{U}_b = f_b(\mathbf{U}_{b-1}), b \in \{1, 2, 3, 4\}, \mathbf{U}_0 = \mathbf{U},
\end{equation}
where the subscript $b$ indicates the $b$-th DPRNN block, and $f(\cdot)$ denotes the mapping function defined by the corresponding block. We adopt bidirectional long short-term memory (BLSTM) to model intra-chunk and inter-chunk dependencies. As illustrated in the right panel of Fig.~\ref{fig:system_overview}, two sub-blocks are used to perform intra- and inter-chunk processing, respectively. Each sub-block comprises of a BLSTM layer and a linear projection layer followed by layer normalization. The intra-chunk sub-block operates on the third dimension of the 3-D representation, and the inter-chunk sub-block on the second dimension. Moreover, a residual connection is used to bypass the input to the output in each sub-block. 

The output $\mathbf{U}_4 \in \mathbb{R}^{N \times S \times R}$ of the last DPRNN block is further processed by a linear projection layer with $N$ units, which is followed by a parametric rectified linear unit (PReLU) function. We perform the overlap-add operation on the resulting 3-D representation at the chunk level, leading to a 2-D representation $Z \in \mathbb{R}^{N \times L}$.

\subsection{Multi-objective learning with transformer blocks}
Once the chunk sequence is modeled by the DPRNN blocks, the learned 2-D representation $Z$ serves as the input to distinct network branches for different metrics, each of which produces an estimate of the corresponding metric score. Each branch consists of a transformer block as illustrated in Fig.~\ref{fig:transformer_block}. Specifically, the 2-D representation $Z$ is first fed into a transformer, which essentially comprises of a multi-head attention module and two linear layers as depicted in Fig.~\ref{fig:transformer_block}. Following~\cite{vaswani2017attention}, the multi-head attention is formulated as
\begin{equation}
\text{MultiHead}(Q, K, V) = \text{Concat}(H_1, \dots, H_h)W^O.
\end{equation}
Each attention head is
\begin{equation}
    H_i = \text{Att}(QW_i^{(Q)}, KW_i^{(K)}, VW_i^{(V)}), i \in \{1, \dots, h\},
\end{equation}
where $W_i^{(Q)} \in \mathbb{R}^{d_q \times d}$, $W_i^{(K)} \in \mathbb{R}^{d_k \times d}$, $W_i^{(V)} \in \mathbb{R}^{d_v \times d}$, $W_i^{(O)} \in \mathbb{R}^{hd_v \times d}$ denote trainable projection weight matrices. We adopt the scaled dot-product attention function, i.e.
\begin{equation}
    \text{Att}(Q', K', V') = \text{Softmax}(\frac{Q'(K')^\top}{\sqrt{d}})V',
\end{equation}
where $Q' = QW_i^{(Q)}$, $K' = KW_i^{(K)}$ and $V' = VW_i^{(V)}$.
Given the single input $Z$, self-attention is performed, where $Q = K = V = Z^\top$ and $d_q = d_k = d_v = N$. The two linear layers in the transformer have $d_1$ and $d$ units, respectively, and the first linear layer is followed by a ReLU function.

We perform auto-pooling on the 2-D representation $X \in \mathbb{R}^{L \times d}$ produced by the transformer, which amounts to a 1-D representation $x \in \mathbb{R}^d$. As outlined in~\cite{mcfee2018adaptive}, the auto-pooling operator can automatically adapt to the characteristics of representations via a learnable parameter $\alpha \in \mathbb{R}$: $x = \sum_{i=1}^L x_i \left(\frac{\text{exp}(\alpha \cdot x_i)}{\sum_{j=1}^L \text{exp}(\alpha \cdot x_j)}\right)$,
where $x_1, \dots, x_L$ is a sequence of features.
The resulting 1-D representation is processed by two consecutive linear layers with $d$ and 1 units, respectively, yielding the scalar output $t$. Note that the first linear layer is followed by a PReLU function. We apply a nonlinear function to the estimated scalar output for STOI and WB-PESQ, which have a value range of roughly $[0, 1]$ and $[1, 4.64]$, respectively. Specifically, we adopt the sigmoid function for both metrics, and additionally perform an affine transformation to accommodate the value range of WB-PESQ, as follows:
\begin{equation}
s = 1 + \sigma(t) \cdot (4.64 - 1),
\end{equation}
where $\sigma$ denotes the sigmoid function.

\begin{table*}
\vspace{-10pt}
\scriptsize
\caption{Investigation of different loss functions and weighting factors. Boldface numbers highlight the best results in each case.}
\centering
\begin{tabular}{c|l|ccc|ccc|ccc|c}
    \hline
    \multirowcell{2}{$\mathcal{L}_s$} & \multicolumn{1}{c|}{\multirow{2}{*}{Weighting Factors}} & \multicolumn{3}{c|}{STOI (\%)} & \multicolumn{3}{c|}{WB-PESQ} & \multicolumn{3}{c|}{SI-SDR (dB)} & \multirowcell{2}{With\\ MTL?} \\
    \cline{3-11}
    & & MAE $\downarrow$ & PCC $\uparrow$ & SRCC $\uparrow$ & MAE $\downarrow$ & PCC $\uparrow$ & SRCC $\uparrow$ & MAE $\downarrow$ & PCC $\uparrow$ & SRCC $\uparrow$  & \\
    \hline
    MSE & $w_1=1, w_2=1, w_3=1$ & 2.606 & 0.929 & 0.919 & 0.193 & 0.925 & 0.939 & 1.269 & 0.973 & 0.969 & No \\
    MAE & $w_1=1, w_2=1, w_3=1$ & 2.442 & 0.934 & 0.928 & 0.175 & 0.938 & 0.950 & 1.163 & 0.977 & 0.974 & No \\
    MAE & $w_1=1, w_2=2, w_3=0.5$ & 2.324 & 0.939 & 0.935 & 0.168 & 0.942 & 0.951 & 1.158 & 0.977 & 0.973 & No \\
    MAE & $w_1=1, w_2=2, w_3=0.5, w_0=0.01$ & 2.310 & 0.936 & 0.934 & 0.165 & 0.944 & 0.954 & 1.129 & 0.978 & 0.975 & Yes \\
    MAE & $w_1=1, w_2=2, w_3=0.5, w_0=0.1$ & 2.182 & 0.942 & 0.943 & 0.157 & 0.949 & 0.956 & 1.010 & 0.982 & 0.980 & Yes \\
    MAE & $w_1=1, w_2=2, w_3=0.5, w_0=1$ & 2.039 & 0.947 & 0.947 & 0.143 & 0.956 & 0.962 & 0.843 & \textbf{0.986} & \textbf{0.985} & Yes \\
    MAE & $w_1=1, w_2=2, w_3=0.5, w_0=1.5$ & 2.018 & 0.949 & 0.947 & 0.143 & 0.957 & 0.962 & 0.841 & \textbf{0.986} & 0.984 & Yes \\
    MAE & $w_1=1, w_2=2, w_3=0.5, w_0=2$ & \textbf{1.994} & \textbf{0.950} & 0.950 & \textbf{0.142} & \textbf{0.958} & \textbf{0.963} & \textbf{0.838} & 0.985 & \textbf{0.985} & Yes \\
    MAE & $w_1=1, w_2=2, w_3=0.5, w_0=2.5$ & 2.035 & \textbf{0.950} & \textbf{0.951} & 0.149 & \textbf{0.958} & \textbf{0.963} & 0.841 & 0.985 & 0.984 & Yes \\
    MAE & $w_1=1, w_2=2, w_3=0.5, w_0=3$ & 2.078 & 0.949 & 0.949 & 0.149 & 0.956 & \textbf{0.963} & 0.849 & \textbf{0.986} & \textbf{0.985} & Yes \\
    MAE & $w_1=1, w_2=2, w_3=0.5, w_0=5$ & 2.001 & 0.949 & 0.950 & \textbf{0.142} & 0.957 & 0.961 & 0.845 & 0.985 & 0.984 & Yes \\
    \hline
\end{tabular}
\label{tab:loss}
\end{table*}

\subsection{Facilitating speech assessment via multi-task learning}
Along with the primary task of metric estimation, we formulate reference signal estimation as a secondary task by introducing an additional output branch, akin to~\cite{yu2021metricnet}. This multi-task learning (MTL) framework can be helpful in two ways. First, intrusive metrics are reference-dependent, and hence providing the underlying reference as a supervisory signal can potentially encourage the shared layers to learn latent representations of the reference signal, which would facilitate improved metric estimation. Second, MTL imposes regularization on the training of shared layers, which we expect can improve generalization capabilities.

\begin{figure}
    \centering
    \includegraphics[width=0.7\columnwidth]{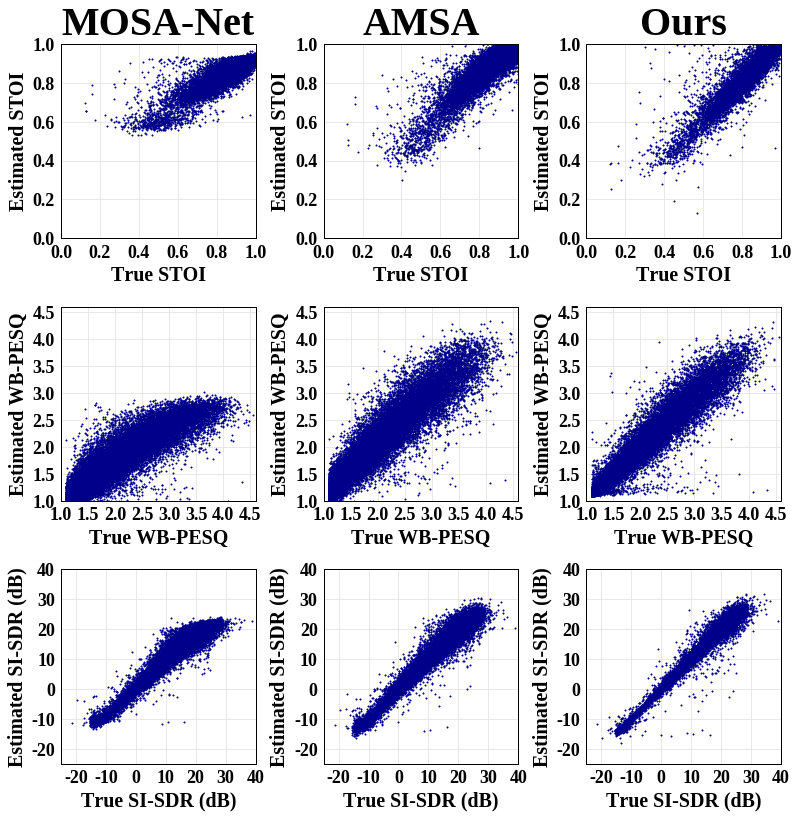}
    \caption{Scatter plots of speech assessment estimates produced by MOSA-Net, AMSA and our model with MTL.}
    \label{fig:scatter_plot}
    \vspace{-10pt}
\end{figure}

As illustrated in the left panel of Fig.~\ref{fig:system_overview}, we use a linear layer with $N$ units as a decoder in the secondary branch. The output 2-D representation is then converted into a time-domain signal through the overlap-add operation at the frame level. All branches are jointly trained to minimize a weighted sum of different losses:
\vspace{-5pt}
\begin{equation}
\mathcal{L} = \sum_{i=1}^3 w_i \cdot \mathcal{L}_s(s_i, \hat{s}_i) + w_0 \cdot \mathcal{L}_z(z, \hat{z}),
\vspace{-5pt}
\end{equation}
where $\hat{s}_i$ and $s_i$ denote the estimated and the corresponding ground-truth metric scores, respectively. $\hat{z}$ and $z$ the estimated and ground-truth reference signals, respectively. The subscript $i = 1, 2, 3$ indicates STOI, WB-PESQ and SI-SDR, respectively. $\mathcal{L}_s$ and $\mathcal{L}_z$ represent the training loss functions for metric and reference signal estimation, respectively, and $w_i,\, i \in \{0,1,2,3\}$ the weighting factor.

\begin{table*}
\scriptsize
\caption{Comparisons of different approaches in terms of MAE, PCC and SRCC. The number of multiply–accumulate operations (MACs) for processing a 5-second audio sample is provided. }
\centering
\begin{tabular}{l|ccc|ccc|ccc|c|c}
    \hline
    \multirow{2}{*}{Approach} & \multicolumn{3}{c|}{STOI (\%)} & \multicolumn{3}{c|}{WB-PESQ} & \multicolumn{3}{c|}{SI-SDR (dB)} & \multirow{2}{*}{\# Params} & \multirow{2}{*}{\# MAC/5s} \\
    \cline{2-10}
    & MAE $\downarrow$ & PCC $\uparrow$ & SRCC $\uparrow$ & MAE $\downarrow$ & PCC $\uparrow$ & SRCC $\uparrow$ & MAE $\downarrow$ & PCC $\uparrow$ & SRCC $\uparrow$ & \\
    \hline
    Quality-Net~\cite{fu2018quality} & - & - & & 0.396 & 0.845 & 0.849 & - & - & - & 0.30~M & 297.30~K \\
    MOSA-Net~\cite{zezario2022deep} & 5.254 & 0.900 & 0.864 & 0.335 & 0.904 & 0.914 & 1.990 & 0.965 & 0.958 & 317.19~M & 94.86~G \\
    AMSA~\cite{dong2020attention} & 3.498 & 0.913 & 0.826 & 0.207 & 0.932 & 0.938 & 1.562 & 0.968 & 0.964 & 2.96~M & 687.61~M \\
    MetricNet~\cite{yu2021metricnet} & - & - & - & 0.182 & 0.938 & 0.947 & - & - & - & 6.61~M & 2.08~G \\
    Ours without MTL & 2.324 & 0.939 & 0.935 & 0.168 & 0.942 & 0.951 & 1.158 & 0.977 & 0.973 & 7.39~M & 40.27~G \\
    Ours with MTL & \textbf{1.994} & \textbf{0.950} & \textbf{0.950} & \textbf{0.142} & \textbf{0.958} & \textbf{0.963} & \textbf{0.838} & \textbf{0.985} & \textbf{0.985} & 7.39~M & 40.27~G \\
    \hline
\end{tabular}
\label{tab:approach_comparison}
\end{table*}

\begin{table*}[ht!]
\scriptsize
\caption{Investigation of multi-objective learning.}
\centering
\begin{tabular}{l|l|ccc|ccc|ccc|c}
    \hline
    \multirow{2}{*}{Approach} & \multicolumn{1}{c|}{\multirow{2}{*}{Weighting Factors}} & \multicolumn{3}{c|}{STOI (\%)} & \multicolumn{3}{c|}{WB-PESQ} & \multicolumn{3}{c|}{SI-SDR (dB)} & \multirowcell{2}{With\\ MTL?} \\
    \cline{3-11}
    & & MAE $\downarrow$ & PCC $\uparrow$ & SRCC $\uparrow$ & MAE $\downarrow$ & PCC $\uparrow$ & SRCC $\uparrow$ & MAE $\downarrow$ & PCC $\uparrow$ & SRCC $\uparrow$ & \\
    \hline
    STOI Alone & \multicolumn{1}{c|}{-} & 2.329 & 0.935 & 0.928 & - & - & - & - & - & - & No \\
    PESQ Alone & \multicolumn{1}{c|}{-} & - & - & - & 0.177 & 0.935 & 0.947 & - & - & - & No \\
    SI-SDR Alone & \multicolumn{1}{c|}{-} & - & - & - & - & - & - & 1.177 & 0.976 & 0.947 & No \\
    Multi-Objective & $w_1=1, w_2=1, w_3=1$ & 2.442 & 0.934 & 0.928 & 0.175 & 0.938 & 0.950 & 1.163 & \textbf{0.977} & \textbf{0.974} & No \\
    Multi-Objective & $w_1=1, w_2=2, w_3=0.5$ & \textbf{2.324} & \textbf{0.939} & \textbf{0.935} & \textbf{0.168} & \textbf{0.942} & \textbf{0.951} & \textbf{1.158} & \textbf{0.977} & 0.973 & No \\
    \hline
\end{tabular}
\label{tab:multi-objective}
\vspace{-5pt}
\end{table*}

\subsection{Experiments}
\subsubsection{Data and Setup}
We use the DNS Challenge 2020 \cite{reddy2020interspeech} dataset in our experiments. Degraded speech are obtained through two primary methods. First, we mix clean speech with additive noise, where the signal-to-noise ratio (SNR) ranges from -15 to 25 dB. Second, we process part of the noisy mixtures randomly with one of three speech enhancement systems. These speech enhancement systems are based on the GCRN architecture~\cite{tan2019learning}, with varying degree of performances due to different configurations. The training, validation, and test set consist of roughly 364500, 14600 and 22800 audio samples, respectively. Due to space constraints, we are unable to show the distribution of PESQ, STOI and SI-SDR in these data, but it covers the value range of 1 to 4.6 for PESQ, 0.25 to 1 for STOI and -18 to 35 dB for SI-SDR. All training signals are truncated to 5 seconds.

For a fair comparison, all models are trained and evaluated on our training and test sets. For our model, we adopt the following configuration for different hyperparameters: $N=256, P=64, R=71, h=4, d=256, d_1 = 1024$. As in~\cite{luo2020dual}, the value of $R$ is selected such that $R \approx \sqrt{2L} \approx S$ for the 5-second training signals. 

\vspace{-0.1in}
\subsubsection{Results and discussion}
We measure the performance of metric estimation using the mean absolute error (MAE), the Pearson correlation coefficient (PCC) and the Spearman’s rank correlation coefficient (SRCC). For both PCC and SRCC, higher scores correspond to better performance. 

Table~\ref{tab:loss} investigates different loss functions and weighting factors. We observe that using the MAE loss as $\mathcal{L}_s$ yields a significantly better performance than using the mean squared error (MSE) loss. In addition, different weighting factor values are compared, revealing that using $w_1=1, w_2=2, w_3=0.5, w_0=2$ with MTL achieves almost the best performance in terms of all the three scores.

Table~\ref{tab:approach_comparison} compares our model with several recent models for deep learning based speech assessment, including Quality-Net~\cite{fu2018quality}, MOSA-Net~\cite{zezario2022deep}, AMSA~\cite{dong2020attention} and MetricNet~\cite{yu2021metricnet}. Note that MOSA-Net and AMSA were developed to estimate multiple metrics simultaneously, while Quality-Net and MetricNet estimates only PESQ. For our model, we use $w_1=1, w_2=2, w_3=0.5$ and $w_0=2$ if MTL is adopted. We observe that our model significantly outperforms all the baselines in terms of MAE, PCC and SRCC. In addition, the performance of our model can be further improved by training with MTL. For example, the MAE, PCC and SRCC for WB-PESQ estimation improves from 0.168, 0.942 and 0.951 to 0.142, 0.958 and 0.963 by using MTL, respectively. We further analyse this in Fig.~\ref{fig:scatter_plot}, where we observe that the data points for our model are more densely distributed near the diagonal relative to MOSA-Net and AMSA, demonstrating that the scores estimated by our model is better correlated with the true scores. 

Our model is based on multi-objective learning, i.e. a single model is learned for multiple metrics simultaneously. We investigate the effect of this strategy over training three different models (only one output branch in Fig.~\ref{fig:system_overview}) with the same architecture to individually estimate STOI, PESQ and SI-SDR. As shown in Table~\ref{tab:multi-objective}, training a single model to estimate multiple metrics simultaneously does not degrade but slightly improve the performance compared with the models that estimate each metric alone. The rationale is that different metrics are correlated with one another and thus each estimation branch regularizes the training of the other branches, which improves the generalization capability of the model. Moreover, such a multi-objective learning approach is computationally more efficient due to the use of shared modules among different objectives.

\vspace{-0.15in}
\section{Conclusions}
\vspace{-0.1in}
We have presented \emph{TorchAudio-Squim}, a system for speech quality and intelligibility assessment. It is released as part of \emph{TorchAudio} in PyTorch which enables easy, accessible uses of deep learning methods to estimate speech quality and intelligibility in a non-intrusive manner. This would not only be useful to various speech systems which require assessment of speech signals but also support research on non-intrusive methods for speech assessment. \emph{TorchAudio-Squim} supports estimation of both subjective and objective speech metrics through novel methods which are shown to outperform prior state-of-the-art methods. Moreover, these models will be continuously developed and improved in future versions of \emph{TorchAudio-Squim}. We intend to extend them to other speech assessment metrics and explore development of computationally more efficient models.

\vfill\pagebreak


\bibliographystyle{IEEEbib}
\bibliography{refs}

\end{document}